\documentclass[conference]{IEEEtran}
\usepackage{graphicx}

\usepackage{cite}

\ifCLASSINFOpdf
\else
\fi
\begin{document}
%
\title{PIMIP: An Open Source Platform for Pathology Information Management and Integration}



%

\author{
\IEEEauthorblockN{Jialun Wu$^{1,\&}$, Anyu Mao$^{1,\&}$, Xinrui Bao$^{2}$, Haichuan Zhang$^{1,3}$, Zeyu Gao$^{1}$, Chunbao Wang$^{4}$,\\ Tieliang Gong$^{1}$, and Chen Li$^{1,*}$}
\IEEEauthorblockA{$^{1}$School of Computer Science and Technology, Xi'an Jiaotong University, Xi'an, China\\
National Engineering Lab for Big Data Analytics, Xi'an Jiaotong University, Xi'an, China} 
\IEEEauthorblockA{$^{2}$School of Automation Science and Engineering, Xi'an Jiaotong University, Xi'an, China}
\IEEEauthorblockA{$^{3}$School of Electrical Engineering and Computer Science, Pennsylvania State University, University Park, PA, USA}
\IEEEauthorblockA{$^{4}$Department of Pathology, the First Affiliated Hospital of Xi'an Jiaotong University, Xi'an, China}
\IEEEauthorblockA{$^{\& }$This two authors contributed equally to this work.\\
Email: maoanyu@stu.xjtu.edu.cn, andylun96@stu.xjtu.edu.cn}
}

\maketitle

\begin{abstract}

Digital pathology plays a crucial role in the development of artificial intelligence in the medical field. The digital pathology platform can make the pathological resources digital and networked, and realize the permanent storage of visual data and the synchronous browsing processing without the limitation of time and space. It has been widely used in various fields of pathology. However, there is still a lack of an open and universal digital pathology platform to assist doctors in the management and analysis of digital pathological sections, as well as the management and structured description of relevant patient information. Most platforms cannot integrate image viewing, annotation and analysis, and text information management. To solve the above problems, we propose a comprehensive and extensible platform, 
PIMIP (\textbf{P}athology \textbf{I}nformation \textbf{M}anagement \& \textbf{I}ntegration \textbf{P}latform).
Our PIMIP has developed the image annotation functions based on the visualization of digital pathological sections. Our annotation functions support multi-user collaborative annotation and multi-device annotation, and realize the automation of some annotation tasks. In the annotation task, we invited a professional pathologist for guidance. We introduce a machine learning module for image analysis. The data we collected included public data from local hospitals and clinical examples. Our platform is more clinical and suitable for clinical use. In addition to image data, we also structured the management and display of text information. So our platform is comprehensive. The platform framework is built in a modular way to support users to add machine learning modules independently, which makes our platform extensible.

\end{abstract}


%
\IEEEpeerreviewmaketitle

\section{Introduction}

Digital pathology has a crucial role in promoting artificial intelligence in the field of medicine. It is widely used in various areas of pathology, such as in pathology morphology related disciplines, teaching and examination of pathology specialist training, pathological subject read communication meeting, the hospital information management, the diagnosis of clinically significant cases, remote diagnosis and consultation, analysis, and communication of scientific research results, image standardization analysis, and statistical analysis and much other work.

The observation and analysis of histopathological images with a large resolution based on the traditional microscope method need to be improved in comprehensiveness and objectivity. File management based on traditional paper mode has low retrieval efficiency and limited storage time. Therefore, an open and universal digital pathology platform is needed to assist doctors in managing and analyzing digital pathological slices and the management and structural description of related patient information. There are few professional and open universal digital slice annotation software, especially the software that supports multi-user collaborative annotation. Moreover, most platforms cannot integrate functions such as image viewing, annotation and analysis, and text information management to form a comprehensive digital pathology platform.

The virtual microscope technology \cite{higgins2015applications} and the OpenSlide \cite{goode2013openslide} library realize the viewing of Whole Slide Images(WSIs). Software such as QuPath \cite{bankhead2017qupath}, and web-based frameworks such as Cancer Digital Slide Archive Project (CDSA \cite{gutman2013cancer}) and its advanced version of Digital Slide Archive Project (DSA \cite{gutman2017digital}) implement image annotation and analysis functions. The proposal of Cytomine \cite{maree2016cytomine} makes it possible for multi-user online annotation. Our platform integrates pathological image viewing, annotation and analysis functions. At the same time, our platform support multi-user collaborative annotation.

The annotation of pathological images is labor-intensive. If it only relies on manual labor will consume a lot of manpower and time. Our platform introduces a part of deep learning-based models for automatic annotation, such as nucleus center annotation, nucleus boundary annotation, etc.

With the rapid development of electronic products, portable office equipment such as mobile phones, tablets, and surfaces have gradually emerged. It is no longer possible to meet the needs of multiple types of equipment by only supporting the annotation function of the computer and the mouse. To solve this problem, our platform is designed to multi-device annotation, such as the cooperation of computer and mouse, the cooperation of tablet and stylus, etc. Additionally, in order to prevent the accidental touch of the touch device from causing the comment disconnection problem, we have introduced a short-time waiting mechanism, so that the break-point comments can be automatically connected into a line.

The Open Histopathological Image(OpenHI) framework \cite{puttapirat2018openhi,puttapirat2019openhi2}  we proposed in 2018 and 2019 uses Simple Linear Iterative Clustering (SLIC \cite{cong2014performance}) superpixel segmentation in image pre-processing, which requires the user to drag and select sub-regions to form region divisions, and the accuracy is not high. To solve the above problems, we propose a comprehensive and extensible platform, PIMIP (\textbf{P}athology \textbf{I}nformation \textbf{M}anagement \& \textbf{I}ntegration \textbf{P}latform).
The PIMIP proposed in this paper uses machine learning models for more accurate area detection. The image analysis function of PIMIP realizes the refinement from region to point, and integrates tumor region segmentation and grading, nucleus detection, grading and segmentation functions. Our platform not only focuses on image resource management, but also realizes the management of patient information associated with slides, and the structuring of descriptive pathological content.

In terms of data, our platform is not limited to The Cancer Genome Atlas (TCGA \cite{tomczak2015cancer}) project which is public, but also introduces clinical data from local hospitals such as the First Affiliated Hospital of Xi'an Jiaotong University in Shaanxi Province and Shaanxi Cancer Hospital. It makes out platform more practical. Additionally, we not only focus on the management and analysis of histopathological images, but also extend our vision to the entire pathology.

The platform proposed in this paper has the following contributions: 
\begin{itemize}
\item We develop a web-based open-source framework to assist doctors in the management of WSIs.
\item We realize the multi-user collaborative annotation and the image analysis based on the deep learning models.
\item We can manage not only the image data but also the text data, and the type of images we have collected expands from histopathology to pathology. We develop a more comprehensive digital pathology platform.
\item Our digital pathology platform focuses more on actual clinical effects. The introduction of clinical data from local hospitals greatly improves the overall practicability of the image analysis models and the platform.
\end{itemize}


\section{Related Work}


The Cancer Genome Atlas (TCGA \cite{tomczak2015cancer}), which provides extensive genomic data in various cancer types, is a large repository of pathological images. It contains large amounts of virtual slides in svs format. The visualization of Whole Slide Images (WSIs) and the integration relevant clinical information play a key role in the digitization of histopathology research.

OpenSlide \cite{goode2013openslide}, an open-source library, have been introduced in 2013. It can handle svs-format files and store virtual slides through pyramid structures. So that it can provide efficient and random access to multi-resolution pathological images. In addition, it provides functions for accessing slide metadata and auxiliary images. Our platform is based on the OpenSeadragon(OSD) library, enabling web-based visualization of WSIs.

The cancer digital slide archive (CDSA \cite{gutman2013cancer}), proposed in 2013, relies on Open Microscope (OMERO) to visualize WSIs. It links pathology with metadata and external data providers and integrates radiology images. Besides, it enables visualization of manual annotations generated with Aperio Imagescope software or machine-generated annotations. CDSA was updated to a more general platform, the Digital Slide Archive(DSA \cite{gutman2017digital}), in 2017. DSA provides hierarchical management of pathological images. Users can annotate images on the platform. The form of annotation includes dot, rectangle, and polygon. Our platform is more comprehensive in image annotation, and can depict precise boundaries at pixel levels for areas.

Annotating the pathological image by a single user is a huge workload task and is prone to errors. Therefore it can't meet the the need for the annotation of pathological images, so there is a need for multi-user collaborative annotation. The Cytomine platform developed in 2016 supports remote collaboration annotations \cite{maree2016cytomine}. Our platform can support multi-user online collaborative annotations, and these annotations on the same slide are integrated and stored in our database. At the same time, in order to reduce manual pressure, our platform introduces models based on deep learning for automatic image annotation. For the rapid development of portable electronic products, our platform supports multiple devices.

The annotation work of pathological images is an important part of pathology research, which is helpful to the analysis of images. Manual analysis is inefficient, but automatic image analysis based on deep learning models helps to extract useful information efficiently from pathological images. The Cytomine platform provides image proofing and quantification algorithms based on machine learning, but the capabilities of recognition are in low level. QuPath, developed in 2017, uses object-oriented hierarchical data models for shallow-to-deep image analysis \cite{bankhead2017qupath}. The DSA developed an algorithm library to integrate various models for image analysis. Our platform not only makes use of image annotations to train deep learning models with better recognition capabilities. Users can also choose appropriate models for image analysis tasks, such as nucleus segmentation, area detection, etc. Our platforms are highly malleable, because the deep learning models can be updated and extended at any time.

We presented a framework in 2018, named Open Histopathological Image (OpenHI \cite{puttapirat2018openhi,puttapirat2019openhi2}) . It achieves the visualization of WSIs, collaborative annotation and the viewing of standardized clinical records. In the pre-processing of pathological images, OpenHI uses SLIC superpixel segmentation to split a WSI into sub-regions. By incorporating superpixel segmentation, users can complete the selection of Region of Interest(ROI). Here, our platform uses deep learning models instead of superpixel segmentation which is of low precision in terms of region segmentation. At the same time, we expand the machine learning module to perform image analysis tasks such as detection, classification, and segmentation. Users can also introduce their own models for image analysis tasks.

\begin{figure*}
    \centering
    \includegraphics[width=180mm]{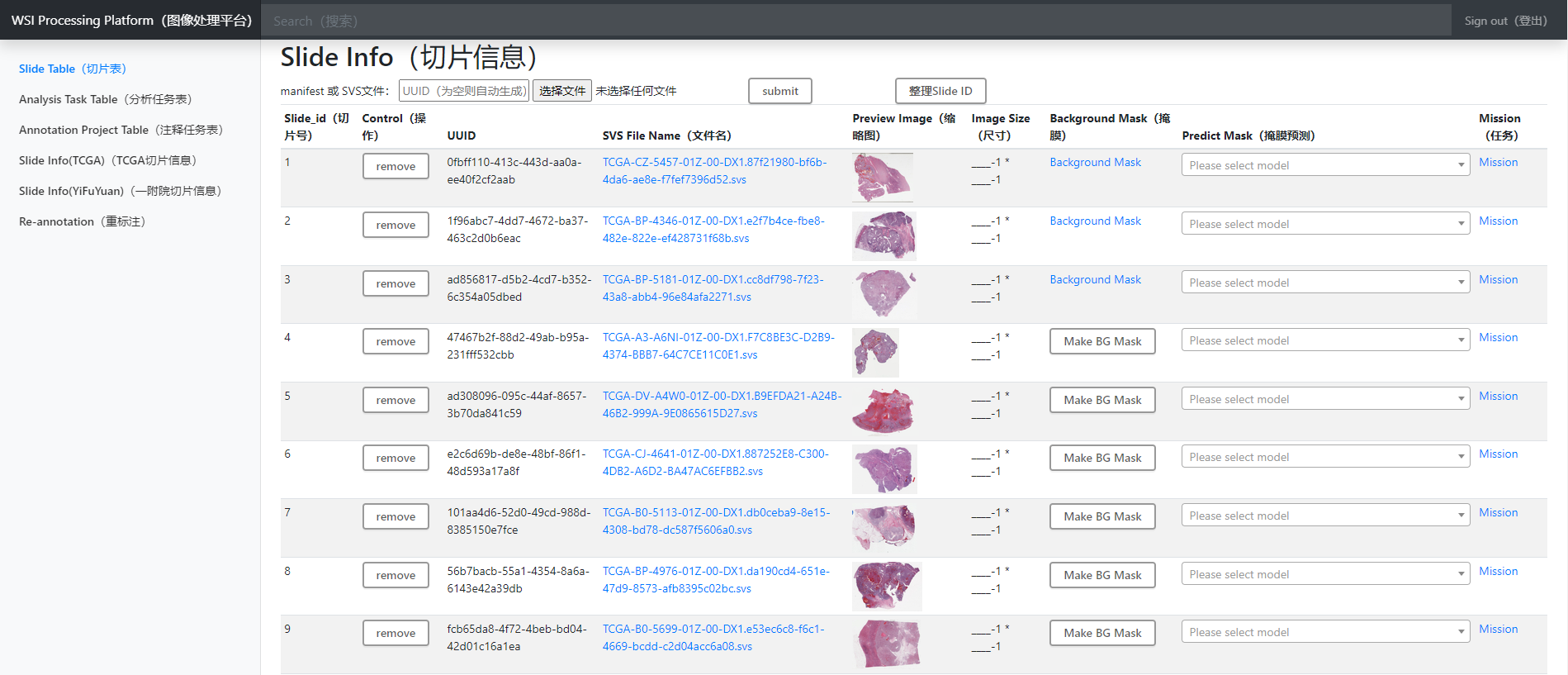}
    \caption{Slide management}
    \label{f1}
\end{figure*}

\section{Material and methods}
In contrast to the OpenHI framework we have proposed, the platform proposed in this paper aims to build a digital pathology platform with data types covering images and texts and data close to actual clinical data. At the same time, the platform's image analysis capabilities are improved by expanding and strengthening the machine learning module.

Here, we will introduce our digital pathology platform by describing the source of data and annotation work, the overall framework and layout of the platform, and the dependence of our platform's functions.

\subsection{The data}


Many histopathological images have been made publicly available by the TCGA project, including the metadata and auxiliary images. Firstly, we downloaded a large amount of WSIs of different cancer types and text data associated with the slides from the TCGA project as the primary data of the platform. These data are used to train machine learning models so that our platform has general analysis capabilities in histopathology images.

In order to improve the performance of the analysis model in clinical practice, we introduce accurate case information data from domestic hospitals, such as kidney and gastric cancer slide images from the First Affiliated Hospital of Xi'an Jiaotong University and tumor slides from Shaanxi Cancer Hospital. In the meantime, to enlarge the scope of pathological images beyond histopathology, we introduce leukemia cell images from Shanghai Children's Medical Center(SCMC).

In the work of image annotation, we invite pathologists from the First Affiliated Hospital for guidance. Pathologists will do a lot of fine-grained annotation work on the data we collect and marking instruction and follow-up inspection work. Firstly, pathologists perform essential annotation work for WSIs based on different image analysis tasks, such as nucleus point labeling, nucleus boundary labeling, and area labeling of tumor slides. After recognizing different areas, the nucleus grading and the cancer sub-type annotation are carried out. Due to the large amount of data collected, pathologists teach us to annotate so that we can annotate pathological images, and the images we labeled will eventually be proofread and checked by pathologists to ensure the correctness of the annotations.

In summary, our image annotation work is guided by pathologists, so that it is professional. The amount of data we have marked covers many diseases. Moreover, our annotations are specific to different image analysis tasks.
\subsection{The framework}

Our platform framework comprises three parts: Web application framework, Graphical user interface(GUI), and machine learning module. It also includes an MYSQL server that stores WSIs data and annotation information.

The OpenHI framework uses SLIC superpixel segmentation technology to perform image pre-processing. This technology divides the image into many irregular sub-regions so that the user can select and merge the sub-regions to complete the area recognition. However, superpixel segmentation does not guarantee that the same type of tissue will be divided into a sub-region, nor can it guarantee that a sub-region contains only one type of tissue, which will cause errors when users merge regions. Image pre-processing by SLIC superpixel segmentation has little help for annotation work. Therefore, our platform does not pre-process the image with superpixel segmentation but uses machine learning models to complete region recognition and segmentation to make image processing more accurate and efficient.

\begin{figure}
    \centering
    \includegraphics[width=80mm]{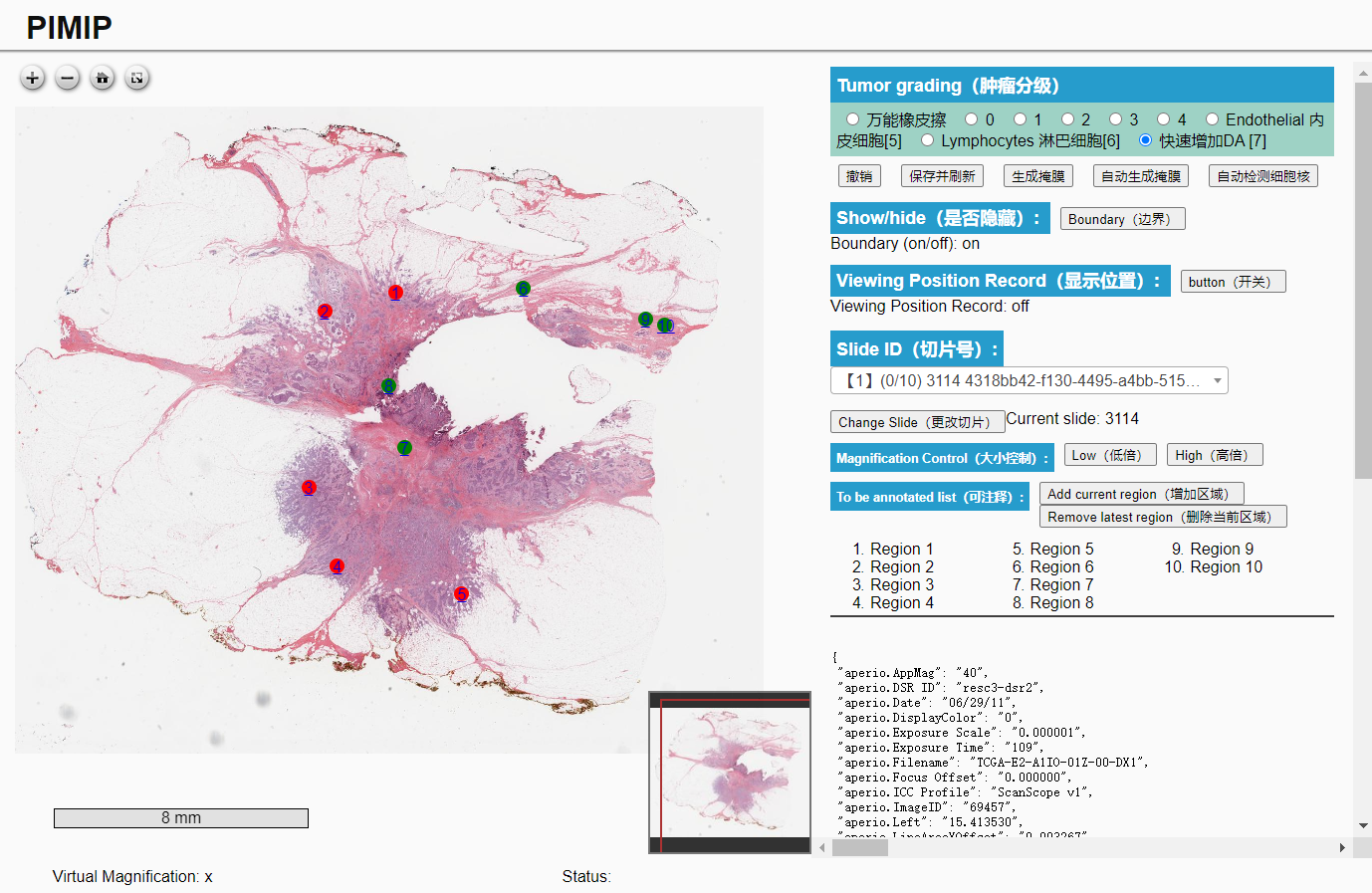}
    \caption{Slide viewer. The viewing of thumbnails combined with related information can be seen in Fig. 1. The viewer in the annotation interface simulates a virtual microscope. Our platform implements a one-key zoom function, and the low/high button can quickly zoom the slide to 20x or 40x, and achieves smooth and random zooming and panning of WSIs based on OSD. On computer devices, users can double-click the mouse to zoom or scroll wheel to zoom. On touch devices, users can zoom by two-finger operation and pan with one-finger operation. }
    \label{f2}
\end{figure}

\subsubsection{Web application framework}


The web server and the browser communicate through the HTTP protocol. Our platform uses the FLASK framework to build web applications, handle HTTP requests and responses, and connect to the database server. Due to the very high resolution and high storage of WSIs, we consider the efficiency issue when processing and only process the request and data transmission of the current observation area, including adding or deleting annotations, annotation visualization.

Considering the particularity of slide visualization and annotation, virtual microscope technology is needed. The OpenSlide library and the OSD algorithm are introduced to store and visualize WSIs in a pyramid structure. That is, each tile has the corresponding enlarged sub-blocks and the shrunken parent blocks. So that users can view a WSI with smooth zooming and panning experience, OSD can also realize the annotation of pathological images by adding custom annotations.

\subsubsection{GUI}

Our platform is a comprehensive digital pathology platform. It is not only for researchers and pathologists but also ordinary learners. Therefore, the graphical user interface must be concise and clear, with high practicability and strong operability.

The home page is a tabbed function browsing interface, and the side tab bar can be understood as a menu bar, which includes the entry of each sub-function interface, such as slide table, mission board, image annotation, associated information view, re-annotation.

In the Slide table, our platform intuitively corresponds to the slide information and the actual image by displaying the thumbnails and related information of WSIs. Click the slide name to view the complete WSI, which is convenient for users to verify. Click the background mask to call the model to recognize the foreground and background, and generate a mask of the actual slide content, which replaces the pre-processing image task with superpixel segmentation, and recognizes the tissue part in WSI. At the same time, it also contains a column of image analysis missions. Click it to view the content and results of the analysis task, which is convenient for researchers to check.

\begin{figure}
    \centering
    \includegraphics[width=80mm]{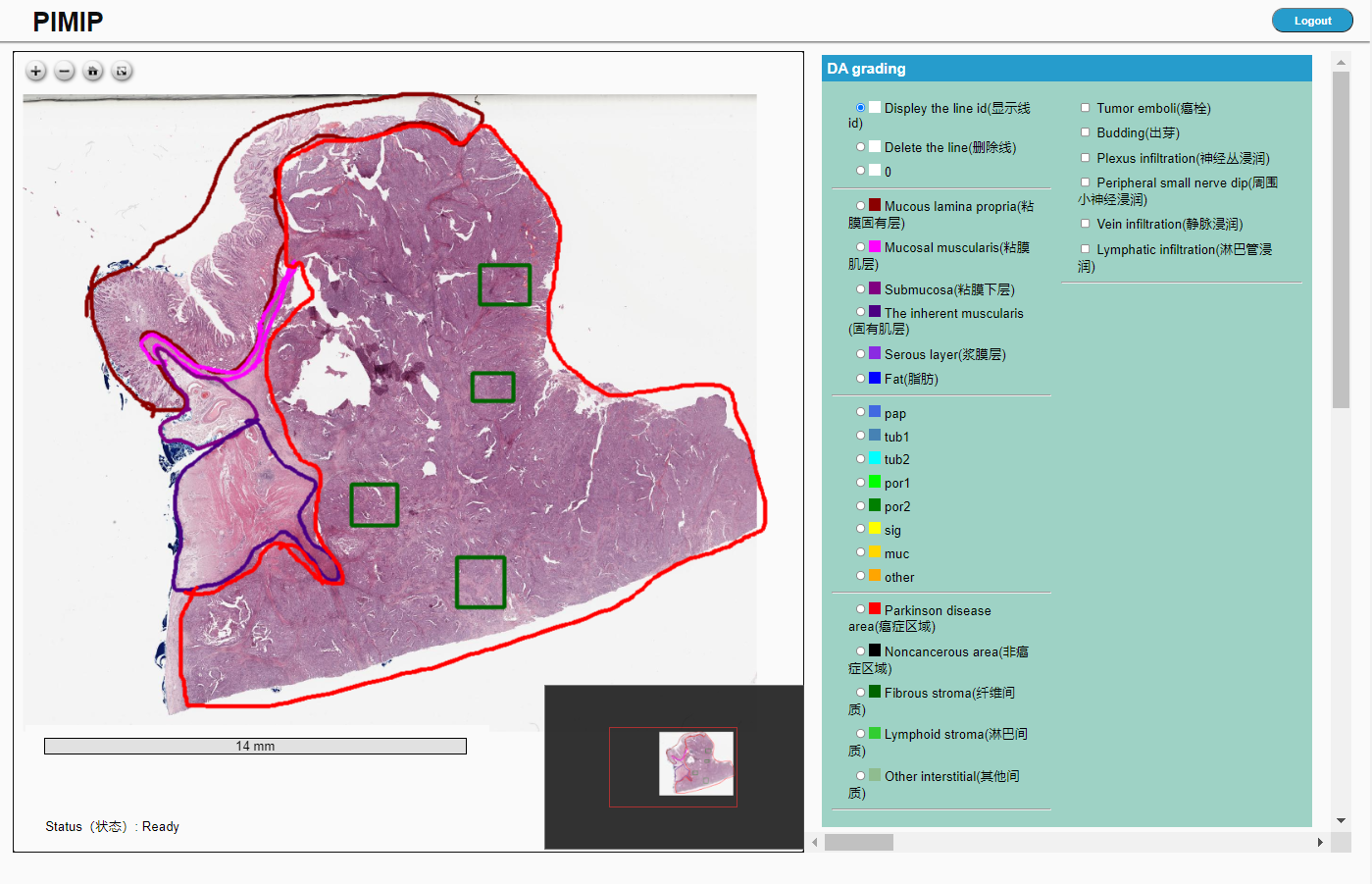}
    \caption{The annotation of rectangle labels and pixel-level precise boundaries}
    \label{f4}
\end{figure}

You can select specific annotation projects in the Annotation Project table, but the corresponding annotation interface has a unified template, convenient for users to master the annotation function. First of all, the WSI visualization part is the viewer. In order to simulate a virtual microscope, we use the low/high button to zoom the image into 20x or 40x quickly. In the annotation task with the computer device, users double-click the mouse or scroll the wheel to zoom the image, and the left mouse button is used to pan the WSI. Touch devices such as tablet phones can zoom in and out through two-finger operation, and panning can be achieved by dragging with one finger. The stylus can smoothly complete annotations on the tablet. The annotation configuration board includes the choice of types of annotation, the slide selection, and the ROI selection. Our platform also has automatic annotation functions based on deep learning. The re-annotation interface includes a correction board for filling or erasing the annotation results.

In the Analysis Task table, our platform allows users to select the image to be analyzed and the analysis model. The results of the analysis will be displayed in the row of the corresponding task.

The slide-related information viewing interface displays a structured pathological report in the form of a form. The user can display the specific information required by column selection and filter the required information through the search function. Click the specific slide name can go to the annotation interface.

\begin{figure*}
    \centering
    \includegraphics[width=180mm]{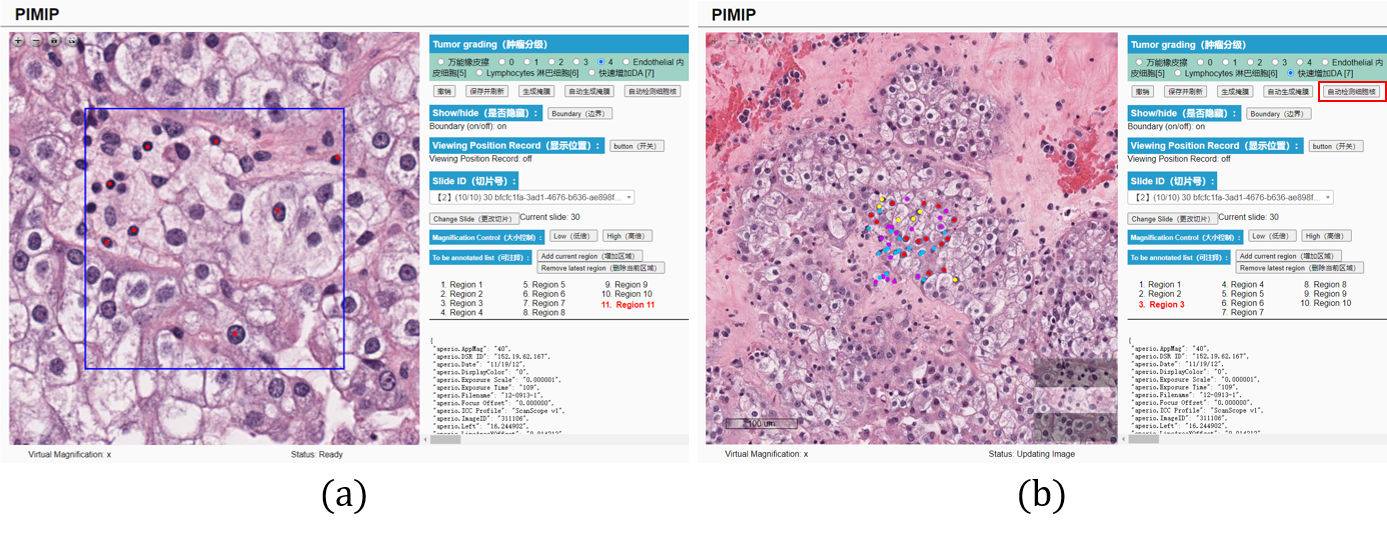}
    \caption{Point labeling. (a) Mark the nuclei manually. (b) Automatically annotate the nuclei using deep-learning models.}
    \label{f3}
\end{figure*}

\subsubsection{Machine learning module}


Our platform uses deep learning models to analyze pathological images, including point detection, nucleus segmentation, region detection, classification, and other functions. We have implemented HoverNet \cite{graham2019hover}, U-Net \cite{2015U}, ResNet \cite{2016Deep}, NuClick \cite{2020NuClick} models. Our team also develops a variety of methods for image analysis, such as \cite{gao2020renal,gao2021instance,gao2021nuclei,shi2019effects,zhang2019classifying}.

At the same time, we use the annotated image data for model training to obtain a higher-precision model and improve image analysis capabilities. The trained model can also be used for automatic annotation in the image annotation function, reducing the pressure on pathologists.

This module is extensible. Users can independently add and select models for analysis tasks. The models used for analysis can also be updated and replaced to ensure that our platform's image analysis keeps pace with the times.

\begin{figure}
    \centering
    \includegraphics[width=80mm]{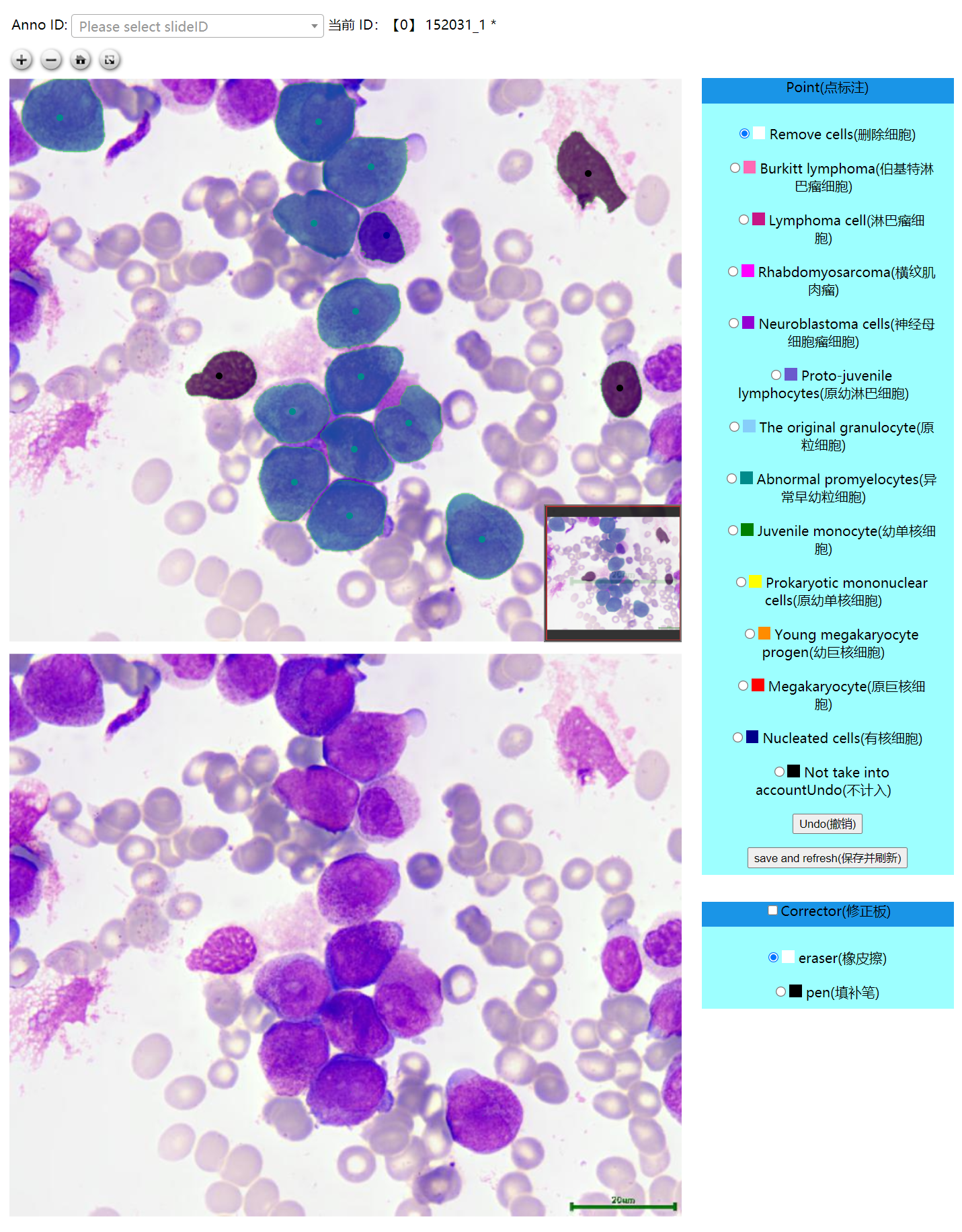}
    \caption{Re-annotation}
    \label{f6}
\end{figure}

\subsubsection{MYSQL server}
The corresponding relationship between the slide ID and the slide file name is stored in the database, conveniently visualizing the WSI file under the correct path. The annotation information of WSIs is stored in the form of a coordinate point array, which is convenient for retrieval, addition, and deletion.

\subsection{The dependence}
The web application of the platform relies on the FLASK framework to build, which is a micro-framework of python with a small volume but rich content. It uses decorators to increase the routes and functions makes it easy to integrate the functions of an interface into one file. In the web-based GUI, the visualization of WSI and the interaction with users rely on OSD. Open-source image processing library such as LibMI \cite{dong2020libmi}, OpenCV \cite{bradski2008learning}, scikit-image \cite{van2014scikit} and Pillow is used to support efficient image reading and writing. The models of the machine learning module are based on the Pytorch framework. The annotation information is stored in the MYSQL server.

\section{Result}
Our platform covers comprehensive functions, including WSI viewer, slide management, image annotation and analysis, and file management. We extend the only browsing function to annotation and analysis, images with structured reports, and histopathology to pathology. We aim to build a comprehensive digital pathology platform.


\begin{figure*}
    \centering
    \includegraphics[width=180mm]{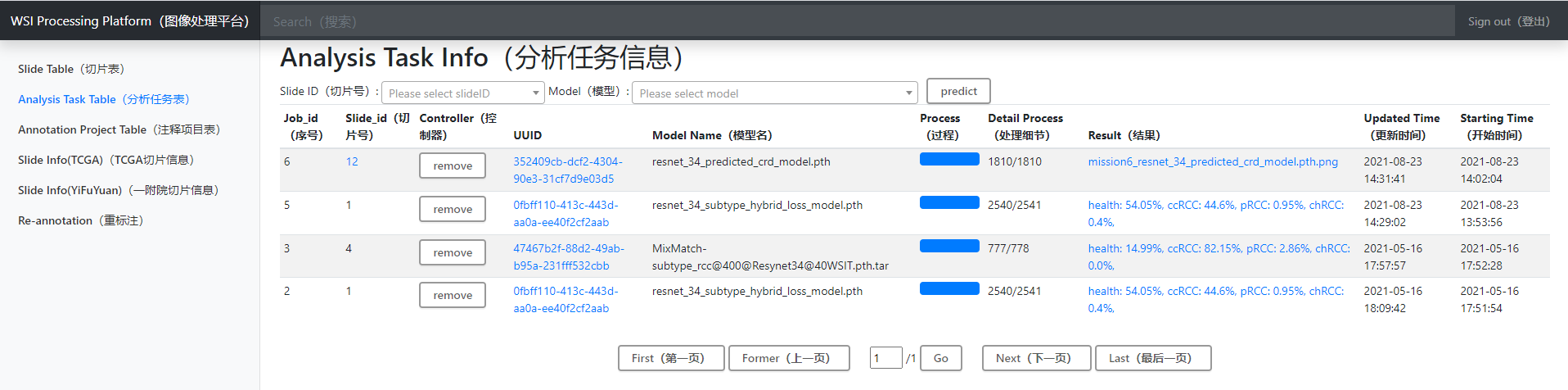}
    \caption{Analysis Table. Users can independently choose images and models to build analysis tasks.}
    \label{f7}
\end{figure*}

\begin{figure}
    \centering
    \includegraphics[width=80mm]{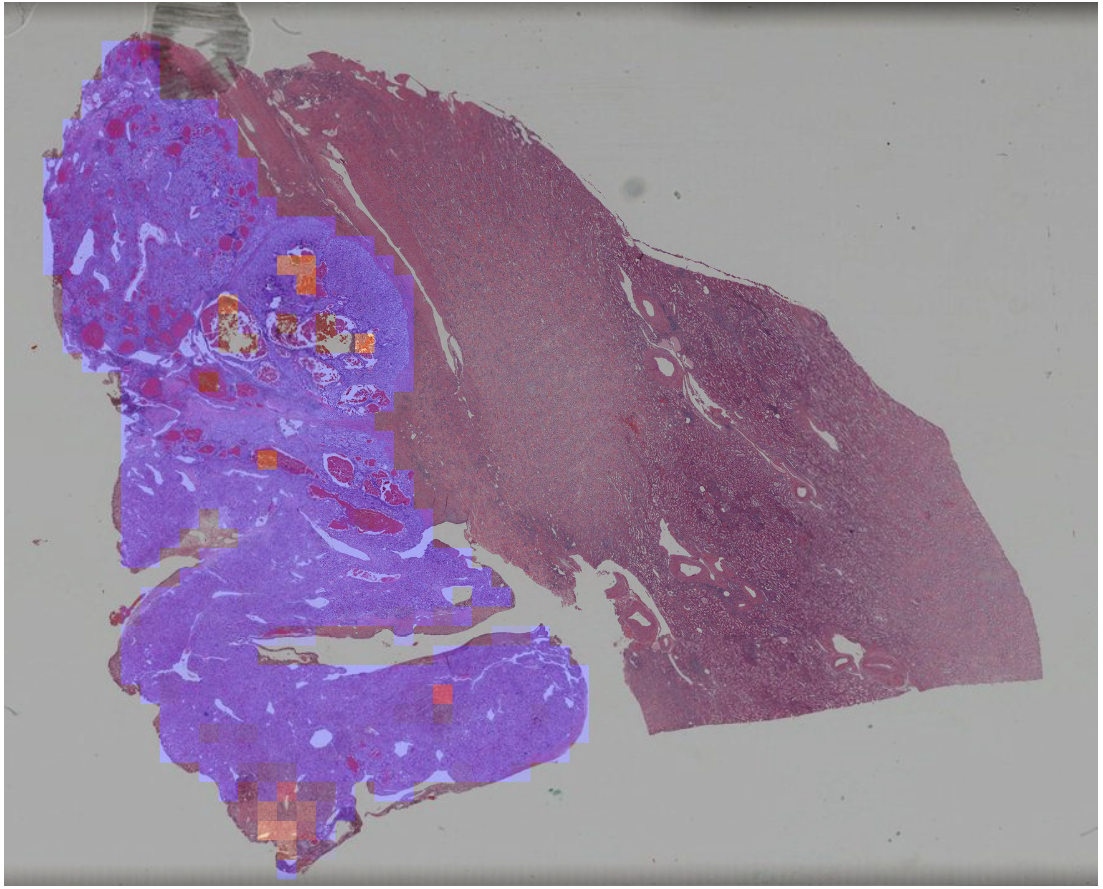}
    \caption{The example of the image analysis in which different types of cancers are highlighted using different colors}
    \label{f8}
\end{figure}

\subsection{Slide management}
\subsubsection{Slide upload and storage}
Our platform has a complete slide management function, and users can upload WSIs in svs format. The slide ID is automatically generated by the platform or named by the user. This ID will be used as the unique identification name of a WSI, which is used to associate with the WSI file name and store it in the MYSQL database. It is also used to create a separate folder when storing annotations, which is convenient for users and platform administrators to retrieve and download data from the resource. The interface for slide upload and storage is shown in Fig. 1.

\subsubsection{Slide viewer}

Firstly, it is the viewing of thumbnails. The large-resolution WSI generates thumbnails, combined with related information. Furthermore, they are displayed together in a form. The viewing of thumbnails of WSIs is shown in Fig. 1.

In the annotation interface, the WSI viewer simulates a virtual microscope. First of all, the platform implements a one-key zoom function, and the low/high button can quickly zoom the slide to 20x or 40x. Secondly, the platform is based on OSD to achieve smooth and random zooming and panning of WSIs. On computer devices, you can double-click the mouse to zoom or scroll wheel to zoom. On touch devices, you can zoom by two-finger operation and pan with one-finger operation. The WSI viewer in the annotation interface is shown in Fig. 2.

\begin{figure*}
    \centering
    \includegraphics[width=180mm]{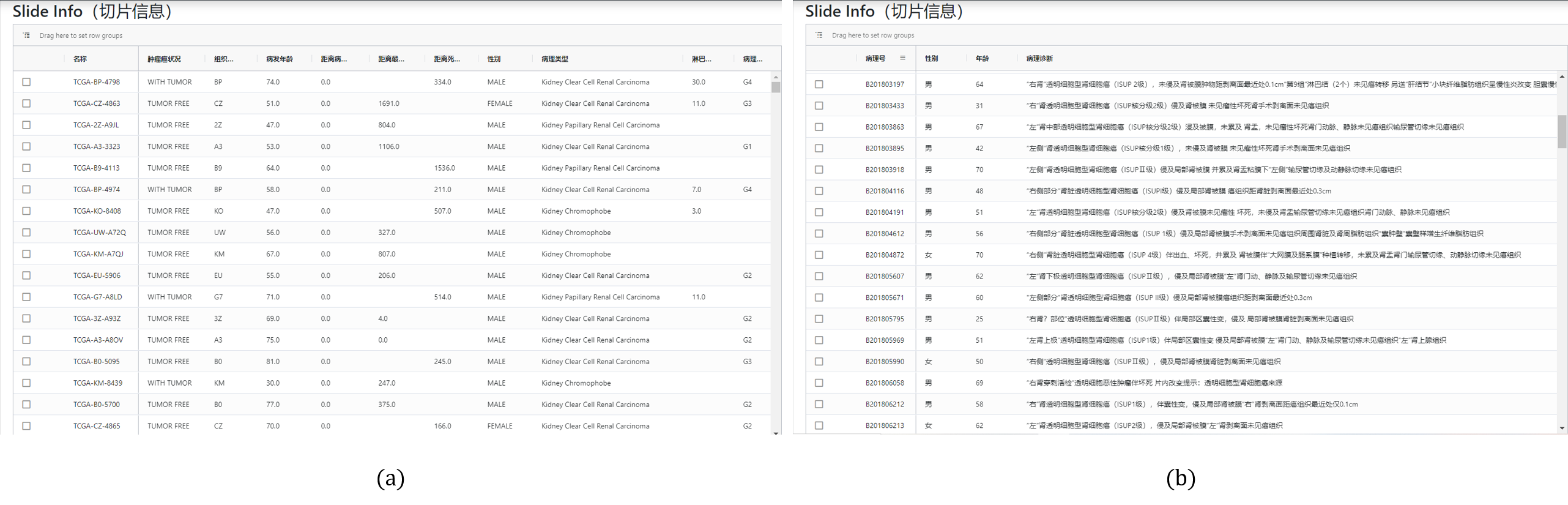}
    \caption{File management. Our platform stores patient information and diagnostic information associated with WSIs, and displays it on the user interface in structured reports, making the descriptive pathology content structured and clear. (a) The data from TCGA. (b) The data from the First Affiliated Hospital of Xi'an Jiaotong University.}
    \label{f9}
\end{figure*}

\subsection{Image annotation}
In our platform, all image annotations can be visualized on the user interface, and there are remedial functions such as erasing, filling, undoing, and clearing for incorrect annotations.
\subsubsection{Point labeling}
Marking the position of the nucleus is the basis annotation of pathological images. Our annotation interface can mark and store the coordinate information of the center of the nucleus by clicking the right mouse button or clicking by the stylus. However, the number of cell nuclei in a WSI is enormous, so point labeling is labor-intensive. Our platform uses a deep learning-based model to realize automatic nucleus point labeling. The manual annotation of nuclei is shown in Fig. 4 (a) and automatic nucleus annotation is shown in Fig. 4 (b).
\subsubsection{Rectangle label}
Users can complete the labeling of the rectangular frame on a WSI, and the platform will record the coordinate information of the four vertices of the rectangular frame and the ID for easy visualization on the user interface. The function of rectangle labeling is shown in Fig. 3.
\subsubsection{Pixel-level precise boundaries}
Users can also draw pixel-level boundaries using a mouse or a stylus to mark irregular boundaries. Our platform will automatically connect the breakpoints generated by the user's mistaken touch during the marking process and outline the closed area. The boundary will be stored in the database as an array of coordinates for easy visualization on the user interface. The function of drawing pixel-level precise boundaries is shown in Fig. 3.
\subsubsection{Re-annotation}
For images that use machine learning models to complete nucleus segmentation, re-annotation increases the center point of the nucleus and calls the semi-supervised model NuClick to add a new nucleus segmentation mask. At the same time, it has the function of erasing or filling the mask to correct the result of nucleus segmentation. The Re-annotation interface is shown in Fig. 5.





\subsection{Image analysis}
Our system supports deep learning modules for analysis of WSIs. There are many deep learning methods for WSIs analysis. \cite{dimitriou2019deep}. Our team also develops a variety of methods for image analysis, such as \cite{gao2020renal,gao2021instance,gao2021nuclei,shi2019effects,zhang2019classifying}. Based on deep learning models, our platform's image analysis is efficient and accurate. At the same time, the annotation information is used to feedback the training models to improve analysis abilities. The analysis table is shown in Fig. 6.

Our platform can distinguish the tissue area from the background and draw the foreground mask to make the image analysis goal clearer. In terms of pathological images of renal cell carcinoma, our platform can realize tasks such as cancer area detection, sub-type classification, and cell nucleus grading.

We use deep learning models such as U-Net \cite{2015U}, W-Net \cite{xia2017w}, HoverNet \cite{graham2019hover}, and NuClick \cite{2020NuClick} to complete the functions of nuclear detection, nuclear segmentation, and nuclear classification. Therefore, our platform construct a full nuclear annotation framework which can complete the nuclei counting and grading tasks to assist pathologists in diagnosing.

We use deep learning models such as U-Net to distinguish between normal and abnormal tissues in tumor slides and classify tumor regions to complete the statistics of the area of the tumor area, the statistics of different tumor types, and assist the pathologist in the diagnosis. An example of tumor regions classification is shown in Fig. 7.

\subsection{File management}
Our platform stores patient information and diagnostic information associated with WSIs, and displays it on the user interface in structured reports, making the descriptive pathology content structured and clear \cite{wu2020structured}. The file management is shown in Fig. 8.

The associated information of the file name and ID of WSIs will be stored in the MYSQL server. The pixel-level precise boundary information will be stored in the database in the form of coordinate arrays. The text or image information generated by the annotation or analysis tasks will be stored in a separate folder named after the slide ID for subsequent export and recall.

\section{Conclusion and Discussion}
In the current situation, there is a lack of professional software to visualize digital pathological images. Even if it is a platform or software that can realize WSI visualization, the function coverage is not comprehensive enough, or it lacks annotation functions or automatic analysis. In order to advance the process of digitization in pathology and solve the problems mentioned above, we have developed a more comprehensive digital pathology platform.
\subsection{Comprehensiveness}
Our platform realizes the visualization of WSI, and our WSI viewer simulates a virtual microscope. Based on visualization, we have implemented the image annotation. Our platform supports multi-user collaborative annotation and multi-device annotation. For labor-intensive annotation tasks, we have developed the automatic annotation function based on deep learning. We also have invited professional pathologists to help in annotation and analysis tasks.

At the same time, we have developed a machine learning module for automatic pathological image analysis. Our machine learning models can also be trained using the annotation information of the WSIs to improve their analysis capabilities.

In addition to image data, our platform organizes and displays the diagnostic information associated with WSIs in a structured manner, covering a more comprehensive range of data types.

In terms of data sources, our platform is not limited to public data but also introduces clinical data from local hospitals, making our platform more effective in clinical applications.
\subsection{Expansibility}
Our platform is extensible. On the one hand, our platform framework is written in modules to facilitate the subsequent addition or deletion of modules or update module functions. On the other hand, our platform allows users to independently add and select machine learning models for analysis in the machine learning module.
\subsection{The Future}
In the future, we hope to update the structured pathology report further, make the descriptive pathology content more structured, easier to interpret by computers, and promote the digitization of pathology. In addition, we hope that the data source of the platform will be closer to clinical practice so that the platform can be more widely used in clinical practice. We will continue to study automatic image annotation models and image analysis models with better performance to improve our platform.


\section*{Acknowledgment}
This work has been supported by National Natural Science Foundation of China (61772409); The consulting research project of the Chinese Academy of Engineering (The Online and Offline Mixed Educational Service System for “The Belt and Road” Training in MOOC China); Project of China Knowledge Centre for Engineering Science and Technology; The innovation team from the Ministry of Education (IRT\_17R86); and the Innovative Research Group of the National Natural Science Foundation of China (61721002). The results shown here are in whole or part based upon data generated by the TCGA Research Network: https://www.cancer.gov/tcga.



%





\end{document}